\documentclass[a4paper,11pt]{article}
\pdfoutput=1 
\usepackage{jheppub} 
\usepackage[T1]{fontenc} 
\usepackage{amsmath,amssymb,relsize,slashed}

\def \beq{\begin{equation}}
\def \eeq{\end{equation}}
\def \bea{\begin{eqnarray}}
\def \eea{\end{eqnarray}}

\def\bm#1{\mbox{\boldmath$#1$\unboldmath}} 

\begin{document} 

\title{Two-loop amplitudes for Higgs plus jet production involving a modified trilinear Higgs coupling}

\author[1]{Martin Gorbahn}
\author[2]{and Ulrich Haisch}

\affiliation[1]{Department of Mathematical Sciences, University of Liverpool, 
\\ L69 7ZL Liverpool, United Kingdom}

\affiliation[2]{Max Planck Institute for Physics, F{\"o}hringer Ring 6, \\ 80805 M{\"u}nchen, Germany}

\emailAdd{Martin.Gorbahn@liverpool.ac.uk}
\emailAdd{haisch@mpp.mpg.de}

\abstract{We calculate the contributions to the two-loop scattering amplitudes  $h \to gg$, $h \to ggg$ and $h \to q \bar q g$ that arise from a modified trilinear Higgs coupling $\lambda$.  Analytic expressions  are obtained by performing an asymptotic expansion near the limit of infinitely heavy top quark. The calculated amplitudes are  necessary to study  the impact of the ${\cal O} (\lambda)$ corrections to the  transverse momentum distributions ($p_{T,h}$) in single-Higgs production at hadron colliders for low and moderate values of $p_{T,h}$.}

\preprint{LTH 1198}


\maketitle
\flushbottom

\section{Introduction}
\label{sec:introduction}

In the Standard Model~(SM) of particle physics the self-interactions of the Higgs field $h$ are given after electroweak~(EW) symmetry breaking by 
\beq \label{eq:VSM}
V \supset \lambda v h^3 + \frac{\chi}{4} \hspace{0.25mm} h^4 \,, \qquad \lambda = \chi = \frac{m_h^2}{2v^2} \,,
\eeq
where $m_h\simeq 125 \, {\rm GeV}$ denotes the Higgs mass and $v \simeq 246 \,{\rm GeV}$ is the  vacuum expectation value~(VEV). One way to  constrain the coefficients $\lambda$ and $\chi$ consists in measuring double-Higgs and triple-Higgs production. Since the cross section for $pp \to 3h$ production is of ${\cal O} (0.1\, {\rm fb})$ at $\sqrt{s} = 14 \, {\rm TeV}$ even the high-luminosity option of the LHC~(HL-LHC) will only allow to set very loose bounds on the Higgs quartic. The prospects to observe double-Higgs production at the HL-LHC is considerably better because the $pp \to hh$ cross section amounts to~${\cal O} (33 \, {\rm fb})$ at the same centre-of-mass energy. Measuring double-Higgs production at the HL-LHC however still remains challenging  and as a result even with the full data set of~$3 \, {\rm ab}^{-1}$ only~${\cal O} (1)$ determinations of the trilinear Higgs coupling $\lambda$ seem possible.

Besides $pp\to hh$, the coefficient $\lambda$ is also subject to indirect  constraints from  processes such as single-Higgs production~\cite{McCullough:2013rea, Gorbahn:2016uoy, Degrassi:2016wml, Bizon:2016wgr, Maltoni:2017ims, DiVita:2017vrr, Maltoni:2018ttu} or EW precision observables~\cite{Degrassi:2017ucl, Kribs:2017znd} since a modified $h^3$ coupling alters these observables at the loop level. In order to describe modifications of the trilinear Higgs coupling in a model-independent fashion, one can employ the SM effective field theory and add dimension-six operators to the SM Lagrangian 
\beq \label{eq:L6}
{\cal L}^{(6)} = \sum_k \frac{\bar c_k}{v^2} \hspace{0.5mm} O_k \,, \qquad O_6 = - \lambda \left |H \right |^6 \,,
\eeq
where $H$ denotes the usual  Higgs doublet. Under the assumption that the effective operator~$O_6$ represents the only relevant modification of the Higgs self-interactions at tree level, one obtains instead of (\ref{eq:VSM}) the result 
\beq \label{eq:VBSM}
V \supset \kappa_\lambda  \hspace{0.25mm} \lambda v h^3 + \kappa_\chi \hspace{0.25mm} \frac{\chi}{4} \hspace{0.25mm}  h^4 \,, \qquad 
\kappa_\lambda = 1 + \bar c_6 \,,  \qquad 
\kappa_\chi = 1 + 6 \hspace{0.25mm} \bar c_6 \,.
\eeq
The second relation allows one to parameterise a modified $h^3$ coupling   via the Wilson coefficient $\bar c_6 = \kappa_\lambda -1$ or equivalent $\kappa_\lambda$. Other operators such as  $O_H = \big (\partial_\mu |H|^2 \big )^2$ or $O_8 =  \left |H \right |^8$ also change this coupling at tree level, but will not be considered here. 

Most of the existing LHC studies that derive constraints on $\lambda$ have assumed that only the $h^3$ vertex is modified while all other Higgs interactions remain SM-like. In~\cite{DiVita:2017eyz} this assumption has been abandoned and ten parameter fits  allowing for modifications $\kappa_\lambda$ have been performed. From these fits one can conclude that standard global Higgs analyses suffer from degeneracies that prevent one from extracting robust bounds on each individual coupling (or Wilson coefficient) once large non-standard $h^3$ interactions are considered. The latter analysis has however also shown that the inclusion of differential measurements in single-Higgs production can help to overcome some of the limitations of a global Higgs-coupling fit that are based on inclusive measurements alone. 

At the differential level the loop-induced effects involving $\bar c_6$ (or $\kappa_\lambda$) are at present  known for vector boson fusion~(VBF), $Vh$~\cite{Degrassi:2016wml, Bizon:2016wgr} as well as  $t \bar t h$~\cite{Degrassi:2016wml} and $t h j$ \cite{Maltoni:2017ims} production, while they have not been calculated in the case of the gluon-fusion channel. The main aim of our work is to close this gap by calculating the relevant two-loop amplitudes for Higgs plus jet production.   The calculation of the  ${\cal O} (\lambda)$ corrections to the two-loop $h \to gg$, $h \to ggg$ and $h \to q \bar q g$ on-shell amplitudes is a multi-scale problem, making it hard but not impossible to obtain exact results. In this article, the computation of the two-loop amplitudes is simplified by performing an asymptotic expansion near the limit of infinitely heavy top quark. The analytic results of our article  will be used  in~\cite{H3pTHiggs}  to obtain predictions for the most relevant differential distributions in Higgs production, such as the transverse momentum of the Higgs~($p_{T,h}$) or jet, in the presence of a modified trilinear Higgs coupling. In the latter article also the prospects of future LHC runs to constrain the Wilson coefficient~$\bar c_6$ using  Higgs plus jets events with low and moderate $p_{T,h}$ will be discussed.
 
This work is organised as follows. In Section~\ref{sec:amplitudes} we discuss the Lorentz structure of the relevant scattering amplitudes and explain how the corresponding form factors can be extracted.  The individual steps of the computations of the form factors are briefly described in~Section~\ref{sec:calculation}.  This section contains a brief discussion of the hard mass procedure that is employed to  obtain the systematic expansions   around the limit of infinitely heavy top quark. Our analytic results for the ${\cal O} (\lambda)$ corrections to the two-loop $h \to gg$, $h \to ggg$ and $h \to q \bar q g$ form factors are presented in  Section~\ref{sec:results}. We conclude in Section~\ref{sec:conclusions}. 

\section{Scattering amplitudes}
\label{sec:amplitudes}

In this section we discuss the parametrisation of the $h \to gg$, $h \to ggg$ and $h \to q \bar q g$ scattering amplitudes in terms of invariant form factors. The extraction of the form factors by means of projection operators is also briefly reviewed. 

\subsection[The $h \to gg$ channel]{The $\bm{h \to gg}$ channel}
\label{sec:hgg}

We start by considering the process $h(p_3) \to g(p_1) + g(p_2)$ and write the corresponding scattering amplitude as 
\beq \label{eq:Agg}
{\cal A}_{gg} =  \delta^{a_1 a_2} \hspace{0.5mm} \epsilon_{1}^{\mu}  (p_1) \epsilon_{2}^\nu (p_2)   \hspace{0.5mm} {\cal A}_{\mu \nu} \,,
\eeq
where $a_{1}$ and $a_{2}$ denote colour indices while $\epsilon_{1}^\mu (p_{1})$ and $\epsilon_{2}^\mu (p_{2})$ are the polarisation vectors of the two final-state gluons. Using Lorentz symmetry and gauge invariance, one can show that the amplitude tensor ${\cal A}_{\mu \nu}$ that appears in (\ref{eq:Agg}) can be expressed in terms of a single form factor~${\cal F}$ in the following way 
\beq \label{eq:Amunu}
{\cal A}_{\mu \nu} = \left ( \eta_{\mu \nu} \hspace{0.5mm} p_1 \cdot p_2 - p_{1 \hspace{0.25mm} \mu} \hspace{0.25mm} p_{2 \hspace{0.25mm} \nu}  \right ) {\cal F} \,,
\eeq
with $\eta_{\mu \nu} = {\rm diag} \left ( 1, -1, -1, -1 \right)$.

The form factor ${\cal F}$ is most conveniently extracted by using a projection procedure. In the case of $h \to gg$ the appropriate projector is (see for instance~\cite{Steinhauser:2002rq})
\beq \label{eq:Pmunu}
P^{\mu \nu} = \frac{1}{\left (d-2\right) \left ( p_1 \cdot p_2 \right)^2} \left ( \eta^{\mu \nu} \hspace{0.25mm} p_1 \cdot p_2 - p_1^\nu \hspace{0.25mm} p_2^\mu - p_1^\mu  \hspace{0.25mm} p_2^\nu \right ) \,,
\eeq
where $d = 4 - 2 \hspace{0.25mm} \epsilon$ denotes the number of space-time dimensions. After applying the projector one can set $p_1^2 = p_2^2 =0$ and $p_1 \cdot p_2 = m_h^2/2$. 

\subsection[The $h \to ggg$ channel]{The $\bm{h \to ggg}$ channel}
\label{sec:hggg}

In the case of the $h(p_4) \to g(p_1) + g(p_2) + g(p_3)$ channel the relevant scattering amplitude can be written as follows
\beq \label{eq:Aggg}
{\cal A}_{ggg} = i \hspace{0mm}  f^{a_1 a_2 a_3} \hspace{0.5mm} \epsilon_{1}^{\mu} (p_1) \epsilon_{2}^\nu (p_2) \epsilon_3^{\lambda} (p_3) \hspace{0.5mm} {\cal A}_{\mu \nu \lambda} \,,
\eeq 
where $f^{a_1 a_2 a_3}$ are the fully anti-symmetric $SU(3)$ structure constants. As before  Lorentz symmetry and gauge invariance restricts the number of possible form factors. In particular, using the transversality conditions $\epsilon_i (p_i) \cdot p_i = 0$ for $i = 1, 2, 3$ and imposing a cyclic gauge fixing condition
\beq \label{eq:cyclic}
\epsilon_1 (p_1) \cdot p_2 = \epsilon_2 (p_2) \cdot p_3 = \epsilon_3 (p_3) \cdot p_1 = 0 \,,
\eeq
the amplitude tensor ${\cal A}_{\mu \nu \lambda}$ can be written in the following way 
\beq \label{eq:Amunula}
\begin{split}
{\cal A}_{\mu \nu \lambda} & = \sum_{n=1}^4 {\cal G}_n  \, T_{n \hspace{0.5mm} \mu \nu \lambda} \,,
\end{split}
\eeq
with  
\beq \label{eq:Tnmunula}
T_{1 \hspace{0.5mm} \mu \nu \lambda} = \eta_{\mu \nu}  \hspace{0.25mm} p_{2 \hspace{0.25mm} \lambda} \,, \quad 
T_{2 \hspace{0.5mm} \mu \nu \lambda} = \eta_{\mu \lambda}  \hspace{0.25mm} p_{1 \hspace{0.25mm} \nu}  \,, \quad 
T_{3 \hspace{0.5mm} \mu \nu \lambda} = \eta_{\nu \lambda}  \hspace{0.25mm} p_{3 \hspace{0.25mm} \mu} \,, \quad 
T_{4 \hspace{0.5mm} \mu \nu \lambda} = p_{3 \hspace{0.25mm} \mu}  \hspace{0.25mm}  p_{1 \hspace{0.25mm} \nu}  \hspace{0.25mm}  p_{2 \hspace{0.25mm} \lambda} \,.
\eeq
The four form factors ${\cal G}_n$ are functions of  the dimensionless ratios 
\beq \label{eq:tauxyz}
\tau = \frac{m_h^2}{m_t^2} \,, \qquad 
x = \frac{s}{m_t^2} \,, \qquad 
y = \frac{t}{m_t^2} \,, \qquad 
z = \frac{u}{m_t^2} \,, 
\eeq
where $m_t \simeq 173 \, {\rm GeV}$ denotes the top-quark mass and 
\beq \label{eq:stu}
s = (p_1 + p_2)^2 \,, \qquad 
t = (p_1 + p_3)^2 \,, \qquad 
u = (p_2 + p_3)^2 \,, 
\eeq
are the partonic Mandelstam variables that fulfil $m_h^2 = s+t+u $. In terms of the variables introduced in (\ref{eq:tauxyz}) the latter relation simply reads $\tau = x+y+z$.

Like in the case of $h \to gg$ the form factors ${\cal G}_n$ can be found by employing an appropriate projection procedure. Following~\cite{Gehrmann:2011aa, Melnikov:2016qoc}, we use 
\beq \label{Pnmunula}
\begin{split}
P_1^{\mu \nu \lambda} & = \frac{1}{d-3} \left [ \frac{t}{s \hspace{0.25mm}  u} \, T_1^{\mu \nu \lambda} - \frac{1}{s \hspace{0.25mm} u} \, T_4^{\mu \nu \lambda}  \right ] \,, \\[2mm]
P_2^{\mu \nu \lambda} & = \frac{1}{d-3} \left [  \frac{u}{s \hspace{0.25mm} t} \, T_2^{\mu \nu \lambda} - \frac{1}{s \hspace{0.25mm} t}  \, T_4^{\mu \nu \lambda}  \right ] \,, \\[2mm]
P_3^{\mu \nu \lambda} & = \frac{1}{d-3} \left [  \frac{s}{t \hspace{0.25mm} u}  \, T_3^{\mu \nu \lambda}  - \frac{1}{t \hspace{0.25mm} u}  \, T_4^{\mu \nu \lambda} \right ] \,, \\[2mm]
P_4^{\mu \nu \lambda} & = \frac{1}{d-3} \left [ -\frac{1}{s \hspace{0.25mm} u} \, T_1^{\mu \nu \lambda}  -\frac{1}{s \hspace{0.25mm} t}  \, T_2^{\mu \nu \lambda}  -\frac{1}{t \hspace{0.25mm} u}  \, T_3^{\mu \nu \lambda} +  \frac{d}{s \hspace{0.25mm} t \hspace{0.25mm} u}  \, T_4^{\mu \nu \lambda}  \right ] \,, 
\end{split}
\eeq
to project out the four different $h \to ggg$ form factors. The tensor structures $T_n^{\mu \nu \lambda}$ have been introduced in (\ref{eq:Tnmunula}). Notice that in order to satisfy (\ref{eq:cyclic}) sums over the external gluon polarisations are taken to be 
\beq \label{eq:sumeps2}
\begin{split}
\sum_{\rm pol.} \big (\epsilon_1^\mu (p_1) \big)^\ast \hspace{0.25mm} \epsilon_1^\nu (p_1) & = -\eta^{\mu \nu} + \frac{p_1^\mu \hspace{0.25mm}  p_2^\nu + p_1^\nu \hspace{0.25mm}  p_2^\mu}{p_1 \cdot p_2} \,, \\[2mm]
\sum_{\rm pol.} \big (\epsilon_2^\mu (p_2) \big)^\ast \hspace{0.25mm} \epsilon_2^\nu (p_2) & = -\eta^{\mu \nu} + \frac{p_2^\mu \hspace{0.25mm} p_3^\nu + p_2^\nu \hspace{0.25mm} p_3^\mu}{p_2 \cdot p_3} \,, \\[2mm]
\sum_{\rm pol.} \big (\epsilon_3^\mu (p_3) \big)^\ast \hspace{0.25mm} \epsilon_3^\nu (p_3) & = -\eta^{\mu \nu} + \frac{p_1^\mu \hspace{0.25mm} p_3^\nu + p_1^\nu \hspace{0.25mm}  p_3^\mu}{p_1 \cdot p_3} \,, 
\end{split}
\eeq
in these projections. 

\subsection[The $h \to \bar q q g$ channel]{The $\bm{h \to q \bar q g}$ channel}
\label{sec:hqqg}

The scattering amplitude describing $h(p_4) \to q (p_1) + \bar q (p_2) + g(p_3)$ takes the form 
\beq \label{eq:Aqqg}
{\cal A}_{q \bar q g} =  t^a_{ij} \hspace{0.5mm} \epsilon_3^{\mu} (p_3) \hspace{0.25mm} {\cal A}_\mu \,,
\eeq
where $t_{ij}^a$ are the colour generators of the fundamental representation of $SU(3)$ with $i$ and~$j$  the colour indices of the quark and the anti-quark, respectively, and $a$ denotes the colour index of the external gluon. The most general ansatz for ${\cal A}_\mu$ consistent with Lorentz symmetry, transversality and parity  involves  two form factors ${\cal H}_m$.  It reads 
\beq \label{eq:Amu}
{\cal A}_\mu = \sum_{m=1}^2 {\cal H}_m \hspace{0.5mm} {T}_{m \hspace{0.5mm} \mu}  \,,
\eeq
with 
\beq \label{eq:T12}
{T}_{1 \hspace{0.5mm} \mu} = \bar u (p_1) \, \big ( \slashed{p}_3 \hspace{0.125mm} p_{2 \hspace{0.25mm} \mu} - p_2 \cdot p_3 \, \gamma_\mu \big) \, v (p_2)   \,, \quad 
{T}_{2 \hspace{0.5mm} \mu}  = \bar u (p_1) \, \big ( \slashed{p}_3 \hspace{0.125mm} p_{1 \hspace{0.25mm} \mu} - p_1 \cdot p_3 \, \gamma_\mu  \big ) \, v (p_2) \,,
\eeq
where $\bar u (p_1)$ and $v(p_2)$ are four-component spinors that describe the external quark fields while $\gamma_\mu$ are the usual Dirac matrices. 

The two form factors entering (\ref{eq:Amu}) can be extracted by applying the following projection operators~\cite{Gehrmann:2011aa, Melnikov:2017pgf}
\beq
\begin{split}
P_1^\mu & = \frac{1}{2 \left ( d-3\right )} \left [ \frac{d-2}{s \hspace{0.25mm} t^2} \left (T_1^\mu \right)^\dagger -  \frac{d-4}{s \hspace{0.25mm} t \hspace{0.25mm} u} \left (T_2^\mu \right)^\dagger \right ] \,, \\[2mm]
P_2^\mu & = \frac{1}{2 \left ( d-3\right )} \left [ \frac{d-2}{s \hspace{0.25mm} u^2} \left (T_2^\mu \right)^\dagger -  \frac{d-4}{s \hspace{0.25mm} t \hspace{0.25mm} u} \left (T_1^\mu \right)^\dagger \right ] \,,
\end{split}
\eeq
with the tensor structures $T_m^{\mu}$ given in (\ref{eq:T12}). After applying these projectors one has to calculate sums over quark, anti-quark and gluon polarisations. For this purpose we employ 
\beq \label{eq:polqqbarg}
\sum_{\rm pol.} u(p_1) \hspace{0.25mm} \bar u(p_1) = \slashed{p}_1 \,, \qquad 
\sum_{\rm pol.} v(p_2) \hspace{0.25mm} \bar v(p_2) = \slashed{p}_2 \,, \qquad 
\sum_{\rm pol.} \big (\epsilon_3^\mu (p_3) \big)^\ast \hspace{0.25mm} \epsilon_3^\nu (p_3)  = -\eta^{\mu \nu} \,.
\eeq
Note that it is allowed to use an unphysical result for the sum over the gluon polarisation since the Dirac structures introduced in (\ref{eq:T12}) are independently transversal. 

\section{Calculation of form factors}
\label{sec:calculation}

\begin{figure}[!t]
\begin{center}
\includegraphics[width=0.975\textwidth]{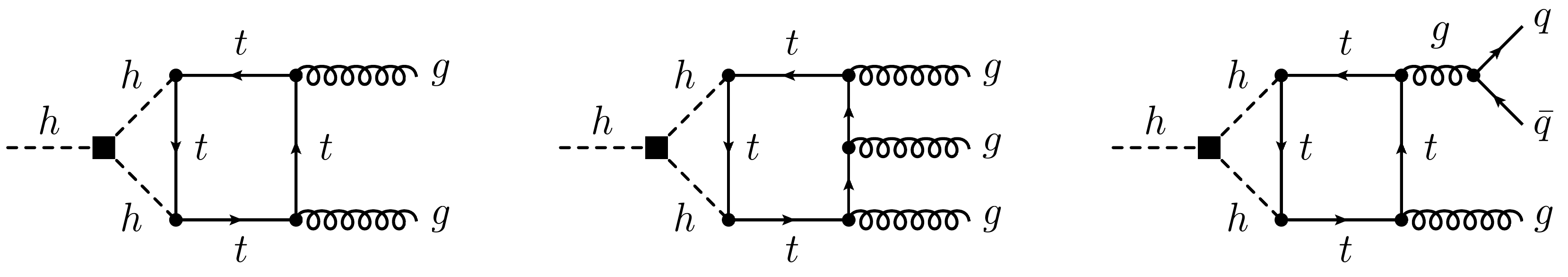}  
\vspace{2mm}
\caption{\label{fig:diagrams}  Examples of two-loop Feynman diagrams with an insertion of an effective trilinear Higgs coupling~(black square) that contribute to the $h \to gg$~(left), $h \to ggg$~(middle) and $h \to q \bar q g$~(right) channel, respectively.}
\end{center}
\end{figure}

Using the projection procedures outlined in the  previous section one can compute each of the $h \to gg$, $h \to ggg$ and $h \to q \bar q g$ form factors separately. Given that the form factors are independent of the external polarisation vectors, all the standard techniques employed in multi-loop computations can be applied. In practice, we proceed as follows. We generate the relevant one-loop and two-loop Feynman diagrams with {\tt FeynArts}~\cite{Hahn:2000kx}. Representative examples of two-loop graphs are shown in Figure~\ref{fig:diagrams}. The actual calculation of the Feynman diagrams is performed in two ways. In the first approach the projection operators are applied diagram by diagram and the resulting loop integrals are then evaluated using the~{\tt FORM} \cite{Vermaseren:2000nd} package~{\tt MATAD}~\cite{Steinhauser:2000ry}. In intermediate steps of the calculation we also make use of the tensor reduction procedures described in~\cite{Tarasov:1995jf,Tarasov:1996br,Tarasov:1997kx} and the~{\tt Mathematica} package~{\tt LiteRed}~\cite{Lee:2013mka} for the reduction of some of the loop integrals. The same techniques have recently also been employed in~\cite{Bizon:2018syu}. The second approach relies entirely on an in-house~{\tt Mathematica} package which calculates the amplitudes algebraically and extracts the form factors at the very end. The agreement of the final results between the two approaches serves as a powerful cross-check of our computations.

The calculation of the  ${\cal O} (\lambda)$ corrections to the two-loop $h \to gg$, $h \to ggg$ and $h \to q \bar q g$ form factors is a multi-scale problem and obtaining exact expressions for the corresponding on-shell amplitudes is therefore notoriously difficult. To simplify the computations we apply the method of asymptotic expansions (for a review see~\cite{Smirnov:2002pj}). Specifically, we work in the limit $m_t^2 \gg m_h^2, s, t, u$ and employ a hard mass procedure to obtain systematic expansions of the relevant two-loop form factors in powers of the ratios $\tau$, $x$, $y$ and $z$ $\big($see~(\ref{eq:tauxyz})$\big)$. Considering the three Feynman diagrams shown in Figure~\ref{fig:diagrams}, it is not difficult to convince oneself that only two types of subgraphs contribute to such an expansion in the case at hand. The first type of contributions arises if the complete diagram is taken to be the subgraph and corresponds to configurations where the external momenta but not the loop momenta are small compared to $m_t$. In this case the asymptotic expansion results in two-loop vacuum integrals with one mass scale that are known analytically since some time~\cite{Avdeev:1994db}. The second type of contributions is obtained by taking only the top-quark loop as a subgraph. Expanding this subgraph in terms of the external as well as the loop momentum running through the Higgs triangle leaves one with one-loop massive vacuum integrals. The corresponding co-subgraphs are one-loop self-energy diagrams that  depend on $m_h$ as well as the external momenta but not on $m_t$. The analytic expressions for such integrals can be found in many textbooks. Combining the two types of contributions and including all diagrams leads to an ultraviolet finite result for the  ${\cal O} (\lambda)$ corrections to the $h \to gg$, $h \to ggg$ and $h \to q \bar q g$ form factors.

\section{Analytic results}
\label{sec:results}

Below we present the analytic results for the ${\cal O} (\lambda)$ corrections to the $h \to gg$, $h \to ggg$ and $h \to q \bar q g$ form factors. Our results have been obtained by the techniques described in the preceding section. 

\subsection[The $h \to g g$ form factor]{The $\bm{h \to g g}$ form factor}
\label{sec:hggffres}

The ${\cal O} (\lambda)$ contribution to the form factor entering (\ref{eq:Amunu}) can be written as follows 
\beq \label{eq:hggcalF}
{\cal F}  = -\frac{\alpha_s}{\pi v} \frac{\lambda}{(4 \pi)^2} \left [ \sum_{p=0}^6 \tau^p  \left ( \frac{Z}{2} \hspace{0.25mm} {\cal  F}^{(p)}_1 +  \bar c_6 \hspace{0.25mm} {\cal F}^{(p)}_2  \right ) \right ] \,.
\eeq
Here $\alpha_s = g_s^2/(4 \pi)$ is the strong coupling constant while 
\beq \label{eq:wavefunction}
Z = \left ( 9 - 2 \sqrt{3} \pi \right ) \bar c_6 \left (\bar c_6 + 2 \right ) \,,
\eeq
denotes the ${\cal O} (\lambda)$ contribution to the Higgs wave function renormalisation constant~\cite{Gorbahn:2016uoy, Degrassi:2016wml}. The one-loop and two-loop coefficients of the asymptotic expansion in $\tau$ of the $h \to gg$ form factor read
\beq \label{eq:hggffs} 
\begin{split}
{\cal  F}^{(0)}_1 & = \frac{1}{3} \,, \qquad {\cal  F}^{(0)}_2  = -L-\frac{\pi }{\sqrt{3}}+\frac{23}{12} \,, \\[2mm]
{\cal  F}^{(1)}_1 & = \frac{7}{360} \,, \qquad {\cal  F}^{(1)}_2  = -\frac{7 \hspace{0.25mm} L}{10}-\frac{7 \hspace{0.25mm} \pi}{20 \sqrt{3}}+\frac{259}{240} \\[2mm]
{\cal  F}^{(2)}_1 & = \frac{1}{504} \,, \qquad {\cal  F}^{(2)}_2  = -\frac{349 \hspace{0.25mm} L}{1008}-\frac{23 \hspace{0.25mm} \pi}{240 \sqrt{3}}+\frac{464419}{1058400} \,, \\[2mm]
{\cal  F}^{(3)}_1 & = \frac{13}{50400} \,, \qquad {\cal  F}^{(3)}_2 = -\frac{1741 \hspace{0.25mm} L}{10800}-\frac{13 \hspace{0.25mm} \pi}{525 \sqrt{3}}+\frac{31795373}{190512000} \,, \\[2mm]
{\cal  F}^{(4)}_1 & = \frac{2}{51975} \,, \qquad {\cal  F}^{(4)}_2  = -\frac{10817 \hspace{0.25mm} L}{138600}-\frac{1789 \hspace{0.25mm} \pi}{277200 \sqrt{3}}+\frac{40370773}{614718720} \,, \\[2mm]
{\cal  F}^{(5)}_1 & = \frac{19}{3027024} \,, \qquad {\cal  F}^{(5)}_2 = -\frac{2798759 \hspace{0.25mm} L}{68796000}-\frac{439357 \hspace{0.25mm} \pi}{252252000 \sqrt{3}}+\frac{2551088981767}{90901530720000} \,, \\[2mm]
{\cal  F}^{(6)}_1 & = \frac{1}{917280} \,, \qquad {\cal  F}^{(6)}_2 = -\frac{1981193 \hspace{0.25mm} L}{86486400}-\frac{991 \hspace{0.25mm} \pi}{2038400 \sqrt{3}}+\frac{277211420687}{20977276320000}\,, \\[2mm]
\end{split}
\eeq 
where we have introduced the shorthand notation $L = \ln \tau$. The coefficients ${\cal  F}^{(p)}_1$ can be easily obtained by a Taylor expansion in $\tau$ from the well-known expression for the top-quark contribution to the on-shell one-loop $h \to gg$ form factor (see for instance~\cite{Gorbahn:2016uoy}).  For $p=0,1,2,3$ the two-loop coefficients ${\cal  F}^{(p)}_2$  agree with~\cite{Degrassi:2016wml}, while the terms with $p=4,5,6$ are presented here for the first time. For the physical value of $\tau \simeq 0.52$ the  terms $\tau^p \hspace{0.25mm} {\cal  F}^{(p)}_2$ with $p=4,5,6$ not included in~\cite{Degrassi:2016wml} amount to an effect of a mere $+0.7\%$, rendering these new higher-order terms in the asymptotic expansion irrelevant for all practical purposes.

\subsection[The $h \to g g g$ form factors]{The $\bm{h \to g g g}$ form factors}
\label{sec:hgggffres}

We write the ${\cal O} (\lambda)$ corrections to the form factors appearing in (\ref{eq:Amunula}) as follows
\beq \label{eq:hgggcalG}
{\cal G}_n  = \, -g_s \, \frac{\alpha_s}{\pi v} \frac{\lambda}{(4 \pi)^2} \left [ \sum_{p=0}^3  \left ( \frac{Z}{2} \hspace{0.25mm} {\cal  G}^{(p)}_{n \hspace{0.25mm} 1} +  \bar c_6 \hspace{0.35mm}  {\cal G}^{(p)}_{n \hspace{0.25mm} 2}   \right ) \right ] \,.
\eeq

In the case of ${\cal G}_1$ the coefficients of the asymptotic expansion of the one-loop contribution proportional to the Higgs wave function renormalisation constant $Z$ read 
\bea \label{eq:Gp11} 
\begin{split}
{\cal  G}^{(0)}_{1 \hspace{0.25mm} 1} & = \frac{ \left (\tau -z \right ) \left (x+z \right )}{3 \hspace{0.25mm} x \hspace{0.25mm} z} \,, \\[2mm]
{\cal  G}^{(1)}_{1 \hspace{0.25mm} 1} & = \frac{7 \hspace{0.25mm} \tau ^2 \left (x+z \right )-\tau  z \left (10 \hspace{0.25mm} x+7 \hspace{0.25mm} z \right )+3 \hspace{0.25mm} x \hspace{0.25mm} z \left (x+z \right )}{360 \hspace{0.25mm} x \hspace{0.25mm} z} \,, \\[2mm]
{\cal  G}^{(2)}_{1 \hspace{0.25mm} 1} & = \frac{10 \hspace{0.25mm} \tau ^3 \left (x+z \right )-\tau ^2 \hspace{0.25mm} z \left (13 \hspace{0.25mm} x+10 \hspace{0.25mm} z \right )+3 \hspace{0.25mm} \tau \hspace{0.25mm} x \hspace{0.25mm} z \left (2 \hspace{0.25mm} x+z \right ) -3 \hspace{0.25mm} x^2 \hspace{0.25mm} z \left (x+z \right )}{5040 \hspace{0.25mm} x \hspace{0.25mm} z} \,, \\[2mm]
{\cal  G}^{(3)}_{1 \hspace{0.25mm} 1} & = \frac{1}{151200  \hspace{0.25mm} x \hspace{0.25mm} z} \, \Big [ \, 39  \hspace{0.25mm}  \tau ^4 \left (x+z \right ) - 3 \hspace{0.25mm}  \tau ^3 \hspace{0.25mm} z \left (19 \hspace{0.25mm} x+13 \hspace{0.25mm} z \right ) + \tau ^2  \hspace{0.25mm} x  \hspace{0.25mm} z  \left ( 74  \hspace{0.25mm}  x+61  \hspace{0.25mm}  z \right ) \\[1mm]
& \hspace{2.5cm} -2  \hspace{0.25mm} \tau   \hspace{0.25mm} x  \hspace{0.25mm} z \left(38  \hspace{0.25mm} x^2+71  \hspace{0.25mm} x  \hspace{0.25mm} z+43  \hspace{0.25mm} z^2\right)+x  \hspace{0.25mm} z \left (x+z \right ) \left(20  \hspace{0.25mm} x^2+43  \hspace{0.25mm} x  \hspace{0.25mm} z+43  \hspace{0.25mm} z^2\right)  \Big ] \,,
\end{split}
\eea
while the two-loop coefficients take the form 
\bea \label{eq:Gp12} 
\begin{split}
{\cal  G}^{(0)}_{1 \hspace{0.25mm} 2} & = 3 \hspace{0.25mm} {\cal  G}^{(0)}_{1 \hspace{0.25mm} 1}  \hspace{0.25mm} {\cal  F}^{(0)}_{2} \,, \\[2mm]
{\cal  G}^{(1)}_{1 \hspace{0.25mm} 2} & =  3 \hspace{0.25mm}  \tau \hspace{0.25mm}  {\cal  G}^{(0)}_{1 \hspace{0.25mm} 1}  \hspace{0.25mm} {\cal  F}^{(1)}_{2} + y \left[ \frac{3}{40} \left ( L + \frac{\pi}{\sqrt{3}} \right ) -\frac{4}{25}\right] \,, \\[2mm]
{\cal  G}^{(2)}_{1 \hspace{0.25mm} 2} & =  3 \hspace{0.25mm}  \tau^2 \hspace{0.25mm} {\cal  G}^{(0)}_{1 \hspace{0.25mm} 1}  \hspace{0.25mm}  {\cal  F}^{(2)}_{2} + y \left [ \frac{2 \hspace{0.25mm} x+3 \left (y+z \right )}{140} \left (  L + \frac{\pi}{2 \sqrt{3}} \right ) -\frac{53903 \hspace{0.25mm} x+ 54421 \left (y+z \right )}{705600}\right ] \,, \hspace{5mm} \\[2mm]
{\cal  G}^{(3)}_{1 \hspace{0.25mm} 2} & =  3 \hspace{0.25mm}  \tau^3 \hspace{0.25mm} {\cal  G}^{(0)}_{1 \hspace{0.25mm} 1}  \hspace{0.25mm}  {\cal  F}^{(3)}_{2} - y \left [ \frac{10463 \hspace{0.25mm} x^2+16575 \hspace{0.25mm} x \left (y+z \right )+5312 \hspace{0.25mm} y^2+12441 \hspace{0.25mm} y \hspace{0.25mm} z+5312 \hspace{0.25mm} z^2}{378000} \, L \right. \\[1mm] 
& \hspace{0.5cm} \left . + \frac{804 \hspace{0.25mm} x^2+1175 \hspace{0.25mm} x \left (y+z \right )+271 \hspace{0.25mm} y^2+1163 \hspace{0.25mm} y \hspace{0.25mm} z+271 \hspace{0.25mm} z^2}{126000} \, \frac{\pi}{\sqrt{3}} \right . \\[1mm]
& \hspace{0.5cm} \left . + \frac{8287709  \hspace{0.25mm} x^2+19944825  \hspace{0.25mm} x \left (y+z \right )+11339441  \hspace{0.25mm} y^2+19028208  \hspace{0.25mm} y  \hspace{0.25mm} z+11339441  \hspace{0.25mm} z^2}{952560000} \, \right ] \,, \hspace{2mm}
\end{split}
\eea 
with the functions ${\cal  F}^{(p)}_{2}$ given earlier in (\ref{eq:hggffs}). 

The form factor ${\cal G}_2$  is obtained from the above expression for ${\cal G}_1$ through the replacements $x \to y$, $y \to z$ and $z \to x$, while in the case of ${\cal G}_3$ the appropriate crossings are $x \to z$, $y \to x$ and $z \to y$.

Our one-loop and two-loop results needed to determine ${\cal G}_4$ are 
\bea \label{eq:Gp41} 
\begin{split}
{\cal  G}^{(0)}_{4 \hspace{0.25mm} 1} & = -\frac{2 \left (x \hspace{0.25mm} y+x \hspace{0.25mm} z+y \hspace{0.25mm} z \right )}{3 \hspace{0.25mm} s \hspace{0.25mm} y \hspace{0.25mm} z} \,, \\[2mm]
{\cal  G}^{(1)}_{4 \hspace{0.25mm} 1} & = -\frac{7 \hspace{0.25mm} x^2 \left (y+z \right )+x \left(7 \hspace{0.25mm} y^2+18 \hspace{0.25mm} y \hspace{0.25mm} z+7 \hspace{0.25mm} z^2\right)+7 \hspace{0.25mm} y \hspace{0.25mm} z \left (y+z \right )}{180 \hspace{0.25mm} s \hspace{0.25mm} y \hspace{0.25mm} z} \,, \\[2mm]
{\cal  G}^{(2)}_{4 \hspace{0.25mm} 1} & = -\frac{\tau  \left(10 \hspace{0.25mm} x^2 \left (y+z \right )+x \left(10 \hspace{0.25mm} y^2+27 \hspace{0.25mm} y \hspace{0.25mm} z+10 \hspace{0.25mm} z^2\right)+10 \hspace{0.25mm} y \hspace{0.25mm} z \left (y+z \right )\right)}{2520 \hspace{0.25mm} s \hspace{0.25mm} y \hspace{0.25mm} z} \,, \\[2mm]
{\cal  G}^{(3)}_{4 \hspace{0.25mm} 1} & =  -\frac{1}{75600 \hspace{0.25mm} s \hspace{0.25mm} y \hspace{0.25mm} z} \, \Big  [ \, 39 \hspace{0.25mm}  x^4 \left (y+z \right )+ 3 \hspace{0.25mm} x^3 \left(39 \hspace{0.25mm}  y^2+ 85 \hspace{0.25mm}  y \hspace{0.25mm}  z+ 39 \hspace{0.25mm}  z^2\right)  \\[1mm] 
& \hspace{2.75cm}   + x^2 \left ( y + z \right )  \left(117  \hspace{0.25mm}  y^2+ 358 \hspace{0.25mm} y \hspace{0.25mm}  z + 117 \hspace{0.25mm}  z^2\right) \\[1mm] 
& \hspace{2.75cm}  + x \left(39 \hspace{0.25mm}  y^4+255 \hspace{0.25mm}  y^3 \hspace{0.25mm}  z+475 \hspace{0.25mm}  y^2 \hspace{0.25mm}  z^2+255 \hspace{0.25mm}  y \hspace{0.25mm}  z^3+39 \hspace{0.25mm}  z^4\right)+39 \hspace{0.25mm}  y \hspace{0.25mm}  z \left (y+z\right )^3 \Big ] \,,
\end{split}
\eea
and 
\bea \label{eq:Gp42} 
\begin{split}
{\cal  G}^{(0)}_{4 \hspace{0.25mm} 2} & = 3 \hspace{0.25mm} {\cal  G}^{(0)}_{4 \hspace{0.25mm} 1}  \hspace{0.25mm} {\cal  F}^{(0)}_{2} \,, \\[2mm]
{\cal  G}^{(1)}_{4 \hspace{0.25mm} 2} & = 3 \hspace{0.25mm}  \tau \hspace{0.25mm}  {\cal  G}^{(0)}_{4 \hspace{0.25mm} 1}  \hspace{0.25mm} {\cal  F}^{(1)}_{2} - \frac{x}{s} \left [  \frac{3}{20} \left ( L + \frac{\pi}{\sqrt{3}} \right ) -\frac{8}{25} \right ] \,, \\[2mm]
{\cal  G}^{(2)}_{4 \hspace{0.25mm} 2} & = 3 \hspace{0.25mm}  \tau^2 \hspace{0.25mm} {\cal  G}^{(0)}_{4 \hspace{0.25mm} 1}  \hspace{0.25mm}  {\cal  F}^{(2)}_{2} - \frac{x}{s} \left [  \frac{3\hspace{0.25mm} \tau}{70} \left ( L + \frac{\pi}{2\sqrt{3}} \right ) - \frac{54421 \hspace{0.25mm} \tau}{352800} \right ] \,, \\[2mm]
{\cal  G}^{(3)}_{4 \hspace{0.25mm} 2} & = 3 \hspace{0.25mm} \tau^3 \hspace{0.25mm} {\cal  G}^{(0)}_{4 \hspace{0.25mm} 1}  \hspace{0.25mm}  {\cal  F}^{(3)}_{2} +\frac{x}{s} \left [  \frac{5312 \hspace{0.25mm} \tau ^2 + 1817 \left (x \left (y+z \right )+y \hspace{0.25mm} z \right)}{189000}   \, L \right.  \\[1mm]
& \phantom{xx} \left. +\frac{271 \hspace{0.25mm}  \tau ^2+621 \left (x \left (y+z \right )+y\hspace{0.25mm}  z \right )}{63000}   \, \frac{\pi}{\sqrt{3}} + \frac{11339441 \hspace{0.25mm} \tau ^2 - 3650674 \left (x \left (y+z \right )+y \hspace{0.25mm} z \right )}{476280000} \right ] \,. \hspace{2mm}
\end{split}
\eea

Notice that all leading-order terms in the asymptotic expansion of the two-loop contribution to the $h \to ggg$ form factors can be written as ${\cal  G}^{(0)}_{n \hspace{0.25mm} 2} \propto {\cal  G}^{(0)}_{n \hspace{0.25mm} 1}  \hspace{0.25mm} {\cal  F}^{(0)}_{2}$. This is an expected feature because these terms can, due to dimensional reasons, only arise from a single effective interaction of the form $h \hspace{0.5mm} G_{\mu \nu}^a G^{a \hspace{0.25mm} \mu \nu}$.  Here $G_{\mu \nu}^a$ denotes the $SU(3)$ field strength tensor. In the heavy top-quark mass limit the same operator however provides the leading contribution to the $h \to gg$ form factor in terms of the function ${\cal  F}^{(0)}_{2}$. The terms ${\cal  G}^{(0)}_{n \hspace{0.25mm} 2}$ hence necessarily have to factorise into the two contributions ${\cal  G}^{(0)}_{n \hspace{0.25mm} 1}$ and ${\cal  F}^{(0)}_{2}$ where the former terms describe the leading corrections in the asymptotic limit to the one-loop $h \to ggg$ form factors.  

\subsection[The $h \to q \bar q g$ form factors]{The $\bm{h \to q \bar q g}$ form factors}
\label{sec:hqqgffres}

The ${\cal O} (\lambda)$ corrections to the form factors  in (\ref{eq:Amu}) can be expressed as 
\beq \label{eq:hqqgcalH}
{\cal H}_m = \, -g_s \, \frac{\alpha_s}{\pi v} \frac{\lambda}{(4 \pi)^2} \left [ \sum_{p=0}^3  \left ( \frac{Z}{2} \hspace{0.25mm} {\cal  H}^{(p)}_{m \hspace{0.25mm} 1} +  \bar c_6 \hspace{0.35mm}  {\cal H}^{(p)}_{m \hspace{0.25mm} 2}   \right ) \right ] \,.
\eeq

In the case of the form factor ${\cal H}_1$ the coefficients of the asymptotic expansion of the one-loop contribution read
\beq \label{eq:Hp11} 
\begin{split}
{\cal  H}^{(0)}_{1 \hspace{0.25mm} 1} & = \frac{1}{3  \hspace{0.25mm} s}  \,, \\[2mm]
{\cal  H}^{(1)}_{1 \hspace{0.25mm} 1} & = \frac{18 \hspace{0.25mm} x+7 \left (y+z \right )}{360 \hspace{0.25mm} s} \,, \\[2mm]
{\cal  H}^{(2)}_{1 \hspace{0.25mm} 1} & = \frac{24 \hspace{0.25mm} x^2+18 \hspace{0.25mm} x \left (y+z \right )+5 \left (y+z \right )^2}{2520 \hspace{0.25mm} s} \,, \\[2mm]
 {\cal  H}^{(3)}_{1 \hspace{0.25mm} 1} & = \frac{100 \hspace{0.25mm} x^3+110 \hspace{0.25mm} x^2 \left (y+z \right )+60 \hspace{0.25mm} x \left (y+z \right )^2+13 \left (y+z \right )^3}{50400 \hspace{0.25mm} s} \, \,,
\end{split}
\eeq 
while the corresponding two-loop contributions are given by 
\beq \label{eq:Hp12} 
\begin{split}
{\cal  H}^{(0)}_{1 \hspace{0.25mm} 2} & = 3 \hspace{0.25mm}  {\cal  H}^{(0)}_{1 \hspace{0.25mm} 1}  \hspace{0.25mm} {\cal  F}^{(0)}_{2} \,, \\[2mm]
{\cal  H}^{(1)}_{1 \hspace{0.25mm} 2} & = 3 \hspace{0.25mm} \tau \hspace{0.25mm}  {\cal  H}^{(0)}_{1 \hspace{0.25mm} 1}  \hspace{0.25mm} {\cal  F}^{(1)}_{2} - \frac{x}{s} \left[  \frac{11 \hspace{0.25mm} L}{45}+\frac{11 \hspace{0.25mm} \pi }{60 \sqrt{3}}-\frac{863}{3600} \right] \,, \\[2mm]
{\cal  H}^{(2)}_{1 \hspace{0.25mm} 2} & =  3 \hspace{0.25mm}  \tau^2 \hspace{0.25mm} {\cal  H}^{(0)}_{1 \hspace{0.25mm} 1}  \hspace{0.25mm}  {\cal  F}^{(2)}_{2} - \frac{x}{s} \left [ x \left(\frac{307 \hspace{0.25mm} L}{1008}+\frac{211 \hspace{0.25mm} \pi }{1680 \sqrt{3}}-\frac{273977}{1058400}\right) \right . \\[1mm] & \left. \hspace{3.5cm} + \left (y + z\right ) \left(\frac{1271 \hspace{0.25mm} L}{5040}+\frac{167 \hspace{0.25mm} \pi
   }{1680 \sqrt{3}}-\frac{266837}{1058400}\right)  \right ] \,, \\[2mm]
{\cal  H}^{(3)}_{1 \hspace{0.25mm} 2} & = 3 \hspace{0.25mm}  \tau^3 \hspace{0.25mm} {\cal  H}^{(0)}_{1 \hspace{0.25mm} 1}  \hspace{0.25mm}  {\cal  F}^{(3)}_{2} - \frac{x}{s} \left [ x^2 \left(\frac{9637  \hspace{0.25mm}  L}{37800}+\frac{503  \hspace{0.25mm}  \pi}{8400 \sqrt{3} }-\frac{4878607}{27216000}\right) \right. \\[1mm] 
& \hspace{3.5cm} + x \left  (y+z \right ) \left(\frac{12407 \hspace{0.25mm} L}{30240}+\frac{4667 \hspace{0.25mm} \pi}{50400 \sqrt{3} }-\frac{16065397}{47628000}\right) \\[1mm] 
& \hspace{3.5cm} + \left. \left (y+z \right )^2 \left(\frac{125863 \hspace{0.25mm} L}{756000}+\frac{9109 \hspace{0.25mm} \pi}{252000 \sqrt{3}}-\frac{1480511}{9720000}\right)  \right ] \, .
\end{split}
\eeq 
The same results hold also for the form factor ${\cal H}_2$. As expected the leading term of the asymptotic expansion of the two-loop pieces again factorises  as ${\cal  H}^{(0)}_{m \hspace{0.25mm} 2}  \propto {\cal  H}^{(0)}_{m \hspace{0.25mm} 1}  \hspace{0.25mm} {\cal  F}^{(0)}_{2}$, since in the infinite top-quark mass limit only the operator $h \hspace{0.5mm} G_{\mu \nu}^a G^{a \hspace{0.25mm} \mu \nu}$ can contribute to the two-loop $h \to q \bar q g$ form factors. 

\section{Conclusions}
\label{sec:conclusions}

In this article we have presented analytic results for the ${\cal O} (\lambda)$ corrections to the two-loop scattering amplitudes $h \to gg$, $h \to ggg$ and $h \to q \bar q g$. These corrections arise in the presence of a modified trilinear Higgs coupling and have been obtained in the form of systematic expansions in the limit $m_t^2 \gg m_h^2, s, t, u$. By a numerical study of the Higgs  transverse momentum $p_{T,h}$~\cite{H3pTHiggs}, we have found that our results show excellent convergence for~$p_{T,h} < m_t$. We thus expected them to provide a reliable approximation to the full on-shell ${\cal O} (\lambda)$ contributions to Higgs plus jet production at low and moderate values of~$p_{T,h}$.  For $p_{T,h} > m_t$ the condition $m_t^2 \gg m_h^2, s, t, u$ is obviously not satisfied and as a result including more terms in the asymptotic  expansion of the form factors (\ref{eq:hgggcalG}) and (\ref{eq:hqqgcalH}) would not improve the calculation of the differential Higgs plus jet production cross section  above the top-quark threshold. In this phase space region a full calculation of the ${\cal O} (\lambda)$ corrections to the on-shell two-loop scattering amplitudes $h \to ggg$ and $h \to q \bar q g$ would be needed to obtain meaningful predictions for Higgs plus jet production. 

With the amplitudes derived in this work, it is now possible to compute the loop-induced effects involving $\bar c_6$ (or $\kappa_\lambda$) to the Higgs boson transverse momentum at low and moderate $p_{T,h}$ not only in the VBF, $pp \to Vh$~\cite{Degrassi:2016wml, Bizon:2016wgr}, $p p \to t \bar t h$~\cite{Degrassi:2016wml} and $p p \to t h j$~\cite{Maltoni:2017ims} channels but also for $pp \to hj$. The phenomenological implications of our  results will be studied elsewhere~\cite{H3pTHiggs}.  In particular, a detailed analysis of the prospects of future LHC runs to constrain the Wilson coefficient $\bar c_6$  using differential information in Higgs plus jets events will be presented there. 

\acknowledgments{The work of MG has been supported by the STFC consolidated grant ST/L000431/1. UH~appreciates  the hospitality and support of the Particle Theory Group at the University of Oxford and the CERN Theoretical Physics Department at various stages of this project.  The research of UH has  also been supported by the Munich Institute for Astro- and Particle Physics (MIAPP) of the DFG cluster of excellence ``Origin and Structure of the Universe''.}


\providecommand{\href}[2]{#2}\begingroup\raggedright\endgroup

\end{document}